\documentclass[aps, twocolumn, pra, showpacs,superscriptaddress]{revtex4}
\usepackage{graphicx}
\usepackage{amsmath}
\usepackage{amsfonts}
\usepackage{amssymb}
\usepackage{epsfig}

\begin{document}

\title{Repeater-assisted Zeno effect in classical stochastic processes}
\author{Shi-Jian Gu}
\email{sjgu@phy.cuhk.edu.hk}
\affiliation{Department of Physics and ITP, The Chinese University of Hong Kong, Hong
Kong, China}
\author{Li-Gang Wang}
\affiliation{Department of Physics and ITP, The Chinese University of Hong Kong, Hong
Kong, China}
\author{Zhi-Guo Wang}
\affiliation{Department of Physics and ITP, The Chinese University of Hong Kong, Hong
Kong, China}
\affiliation{Department of Physics, Tongji University, Shanghai, 200092, China}
\author{Hai-Qing Lin}
\affiliation{Department of Physics and ITP, The Chinese University of Hong Kong, Hong
Kong, China}

\begin{abstract}
As a classical state, for instance a digitized image, is transferred through
a classical channel, it decays inevitably with the distance due to the
surroundings' interferences. However, if there are enough number of
repeaters, which can both check and recover the state's information
continuously, the state's decay rate will be significantly suppressed, then
a classical Zeno effect might occur. Such a physical process is purely
classical and without any interferences of living beings, therefore, it
manifests that the Zeno effect is no longer a patent of quantum mechanics,
but does exist in classical stochastic processes.
\end{abstract}

\date{\today }
\maketitle



\section{Introduction}

Quantum Zeno effect, proposed by Misra and Sudarshan \cite{BMisra77} in
1977, is believed to exist uniquely in the quantum world due to informatical
description of quantum states and the projective measurement in quantum
mechanics (for a review, see Ref. \cite{KKoshinoPhysRep}). While its
classical correspondences, such as the Fletcher's paradox, becomes an
antinomy in Newtonian mechanics. The basic reason, from our judgement, is
the absence of the projective measurement in classical physics. However,
though Newtonian mechanics excludes the projective measurement, the later
does exist in the classical world. In a recent work on this issue \cite%
{SJGuMario}, one of us (Gu) touched the possibility of the classical Zeno
and anti-Zeno effects. Gu used a scenario of Super Mario's prison break to
show that a \textquotedblleft classical state" might not decay if it is
observed continuously. Nevertheless, the scenario involves too much
subjective freedoms coming from Super Mario's intelligent feedback, such
that it is, though reasonable in everyday life, far beyond the objective
laws of the classical world. This consideration leads to the main motivation
of the present work to seek a possibility of the classical Zeno effect
without involving any subjective judgement.

\begin{figure}[tbp]
\includegraphics[width=8cm]{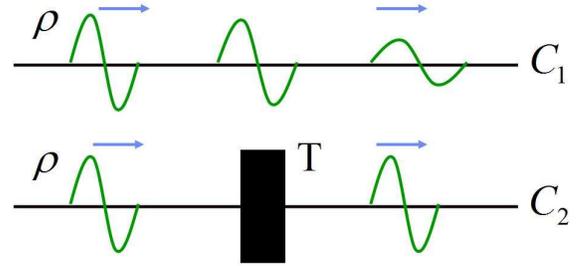}
\caption{A sketch of transferring a classical state $\protect\rho$ through a
classical channel. UP: $\protect\rho$ is transferred through the channel
without any repeater and decays with the distance. LOW: A repeater T is
installed to check the completeness of the state, then $\protect\rho$ might
decay more slowly. }
\label{sketch.eps}
\end{figure}

We consider here such a scenario of transferring a classical state, for
instance a digitized image, through a classical channel with a noise
surrounding (See Fig. \ref{sketch.eps}). The state, if there is no assisted
equipment, decays inevitably with the distance due to the surroundings'
interferences, and finally loses all information. In order to suppress the
decay rate or even ensure zero decay, we need to check the information of
state within a certain distance. Instead of the intelligent judgement made
by Super Mario \cite{SJGuMario}, here we use a machine (repeater), which has
nothing to do with living beings, to examine and try to recover the state.
We show that if there are enough number of repeaters, the transferred state
then might decay much more slowly, or even never decay, then a classical
Zeno effect occurs.

This work is organized as follows. In section \ref{sec:class}, we introduce
the basic formulas for describing the classical transportation. In section %
\ref{sec:example}, we take the emblem of the Chinese University of Hong Kong
(CUHK) as an explicit example to show how the classical Zeno effect occurs
during a classical transportation. A further discussion and a brief summary
are given in section \ref{sec:dis} and \ref{sec:sum} respectively.

\section{Classical transportation}

\label{sec:class}

To begin with, we consider a classical $N$-bit state describe by $\rho
=\{\sigma _{j}\}$ with $\sigma _{j}=\pm 1$, which in computer science can be
saved physically by for instance in a flash disk or a perforated paper tape.
Now we want to transfer the state through a classical channel, which is subject
to a noise background of temperature $T$. During the transferring process, the
state carried by each bit is protected via an energy gap $\Delta E$. So if the
temperature is zero, there is no background inference, and the state can be
transferred over an infinite distance without loss of information. However, the
thermal fluctuation of the surrounding might flip the state of bit. Once the
bit is flipped one or more times, we assume it loses the correlation to the
original global state. The transition probability of each bit through a unit
distance $a$ is determined by
\begin{equation}
P(\sigma _{j}\rightarrow \overline{\sigma }_{j})=e^{-\Delta E/T},
\end{equation}%
where $\overline{\sigma }_{j}$ denotes uncorrelated state. Then if the bit
is transferred over a distance of $L$ (in unit of $a$), the surviving
probability becomes
\begin{eqnarray}
P_{S}(\sigma _{j}) &=&(1-e^{-\Delta E/T})^{L} \\
&=&\exp (-L\Delta E/T^{\prime }),  \notag
\end{eqnarray}%
where $T^{\prime }=-\Delta E/\ln (1-e^{-\Delta E/T})$. Therefore, the
classical state will decay exponentially to a complete random state with the
transferred distance.

\begin{figure*}[tbp]
\includegraphics[width=15cm]{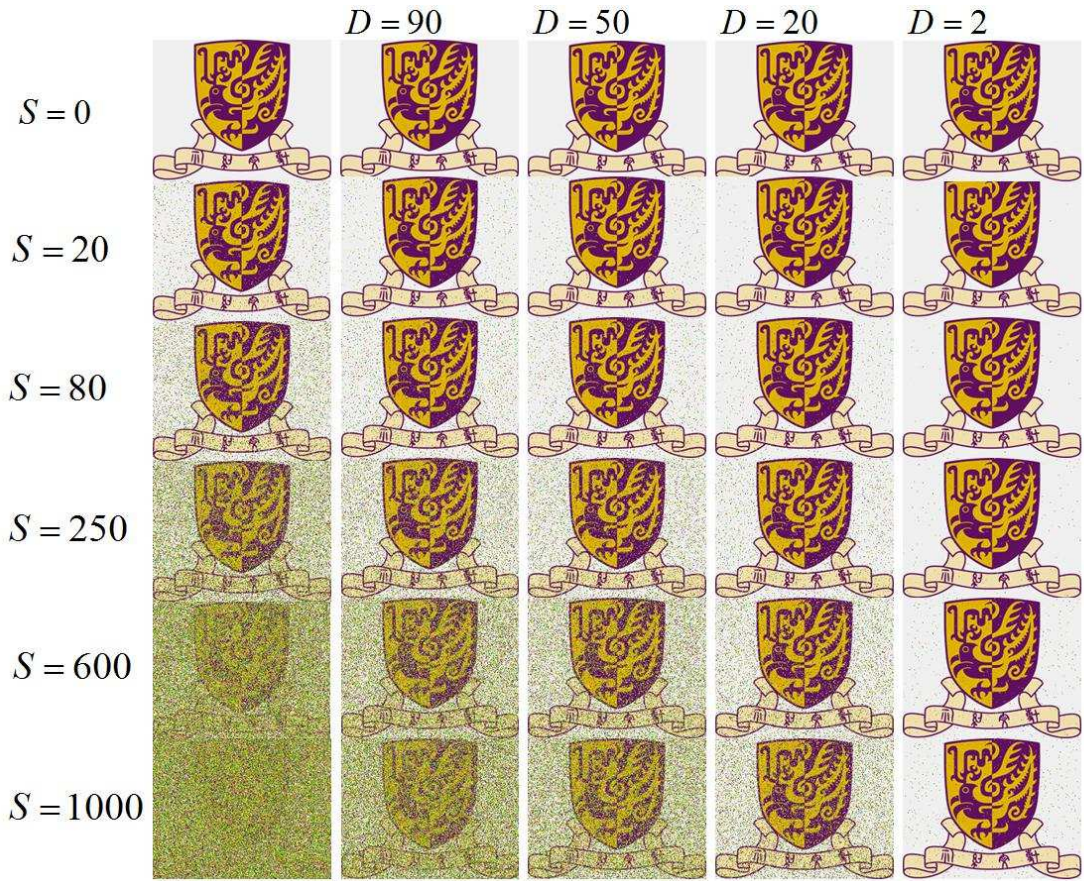}
\caption{The CUHK emblem is transferred trough a classical channel. Here $%
\protect\beta =6$ and S is in unit of $20000$ steps. The leftmost column
denotes the transferring without any repeaters; while other columns
correspond to repeater-assisted cases for various inter-distances of $D=90,
50, 20, 2$ respectively. }
\label{cuhk.eps}
\end{figure*}

Explicitly, in the basis of $\{\sigma _{j},\overline{\sigma }_{j}\}$, the
initial state of the bit is
\begin{equation}
\rho =\left(
\begin{array}{cc}
1 &  \\
& 0%
\end{array}%
\right) ,
\end{equation}%
while the final state becomes%
\begin{equation}
\rho ^{\prime }=\left(
\begin{array}{cc}
\exp (-L\Delta E/T^{\prime }) &  \\
& 1-\exp (-L\Delta E/T^{\prime })%
\end{array}%
\right) .
\end{equation}%
The distance between these two states can be measured by the fidelity
\begin{equation}
F=\text{tr}\sqrt{\rho ^{1/2}\rho ^{\prime }\rho ^{1/2}}=\exp (-L\Delta
E/2T^{\prime }).  \notag
\end{equation}%
Then at the very beginning, i.e. small $L$
\begin{equation}
F=1-\frac{L\Delta E}{2T}+\cdots ,  \notag
\end{equation}%
which decays algebraically. We notice that the leading term of the fidelity
here is linear instead of quadratic in quantum fidelity. Nevertheless, the
algebraic decay of the fidelity ensure that we have chance to correct the
lost information via a repeater.

\section{Example}

\label{sec:example}

In this section, we give a simple example to illustrate how such a process
can be realized in practice. We assume that Alice wants to transfer a
picture, i.e, the CUHK emblem, to Bob who is at a long distance via a
classical channel. The emblem is prepared in a BMP file with $252\times 200$
resolution and 256 colors. For simplicity, We assume Alice and Bob have
already known all other structural information, such as 256 elemental colors
and the size of the picture, except the image data, which can be expressed
as a $252\times 200$ matrix $M$
\begin{equation}
M=\left(
\begin{array}{cccccc}
M_{1}^{1} & M_{2}^{1} &  & \cdots  &  & M_{252}^{1} \\
M_{1}^{2} & M_{2}^{2} &  &  &  &  \\
&  & \ddots  &  &  &  \\
\vdots  &  &  & M_{i}^{j} &  & \vdots  \\
&  &  &  & \ddots  &  \\
M_{1}^{200} &  &  & \cdots  &  & M_{252}^{200}%
\end{array}%
\right) ,  \notag
\end{equation}%
with each element $M_{i}^{j}$ ($i\in \lbrack 1,252],j\in \lbrack 1,200]$)
being an 8-bit integer. During the transportation, each element of the
matrix $M$ is subject to environmental interference. One of elements might
be changed to a random 8-bit integer with the probability $\exp (-\beta )$
with $\beta =-\Delta E/T$ over the unit distance $a$. Clearly, the lower the
temperature, the more stable the data. Meanwhile, as the transferred
distance increase, the emblem's information will inevitably lose. We use the
classical Monte Carlo method to simulate such a process. The leftmost column
of Fig. \ref{cuhk.eps} shows the emblem at the various distance. We can see
that the figure will finally be blurred.

In order to suppress the decay rate, Alice and Bob need to install repeaters
on the channel they use. The repeater's first function is to check the
integrability of data. Secondly, if possible, it can partially recover the
integrability or require the previous repeater to resend the data. For this
purpose, they add row-column check code to the end of the matrix elements.
\begin{equation}
\begin{array}{ccc}
M= & \left(
\begin{array}{cccccc}
M_{1}^{1} & M_{2}^{1} &  & \cdots &  & M_{252}^{1} \\
M_{1}^{2} & M_{2}^{2} &  &  &  &  \\
&  & \ddots &  &  &  \\
\vdots &  &  & M_{i}^{j} &  & \vdots \\
&  &  &  & \ddots &  \\
M_{1}^{200} &  &  & \cdots &  & M_{252}^{200}%
\end{array}%
\right) & \left(
\begin{array}{c}
M_{{}}^{1} \\
M_{{}}^{2} \\
\vdots \\
M_{{}}^{j} \\
\vdots \\
M_{{}}^{200}%
\end{array}%
\right) \\
& \left(
\begin{array}{cccccc}
\text{ }M_{1}\text{ } & \text{ }M_{2}\text{ } & \text{ \ } & M_{i}\text{ } &
\text{ \ } & \text{ \ }M_{252}\text{ }%
\end{array}%
\right) &
\end{array}%
\end{equation}%
The row and column check codes are defined as%
\begin{equation}
M_{i}=\sum_{j}M_{i}^{j},M^{j}=\sum_{i}M_{i}^{j}  \label{eq:checkcode}
\end{equation}%
respectively. So if one of elements $M_{i}^{j}$ or $M_{i}(M^{j})$ is
changed, the corresponding row and column check code do not match. Since the
Eq. (\ref{eq:checkcode}) are simple summation, in this case, the repeater is
able to recover the lost information, that is%
\begin{equation}
M_{i}^{j}=M_{i}-\sum_{l\neq j}M_{i}^{l}
\end{equation}%
or%
\begin{equation}
M_{i}^{j}=M^{j}-\sum_{l\neq i}M_{l}^{j}.
\end{equation}
If one of elements in $M_{i}(M^{j})$ is lost, the repeater recalculate $%
M_{i}(M^{j})$ from the matrix $M$. \textit{For simplicity}, if there are two
or more elements are lost, we assume that the repeater is neither able to
recover the integrability, nor requires the previous repeater to resend the
picture again, but recalculate Eq. (\ref{eq:checkcode}) to ensure the
integrability of data for next repeater. We show our results in Fig. \ref%
{cuhk.eps}. In the second column, the distance between two repeaters is $%
D=90a$. The emblem keep much more information than the first column.
Moreover, if we add more repeaters to the channel, the decay rate will be
suppressed further. Especially if $D=2a$, the emblem at $L=2\times 10^{9}a$
is almost the same as the original one. We then call such a phenomenon as a
repeater-assisted classical Zeno effect in random processes.

\begin{figure}[tbp]
\includegraphics[width=8cm]{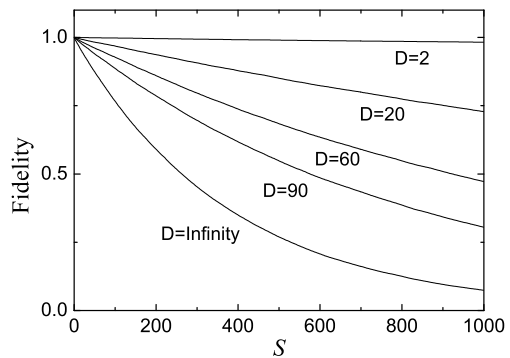}
\caption{The fidelity between the initial state and the state during
transportation as a function of $S$ (in unit of 20000 steps) for various $D$%
. }
\label{fidelity.eps}
\end{figure}

Precisely, we can introduce the fidelity to describe the honesty of the
state to the original one. In this case, the fidelity can be defined as
\begin{equation}
F=\frac{\text{number of unchanged pixel}}{\text{total number of pixel}}.
\end{equation}
The results are shown in Fig. \ref{fidelity.eps}, from which we can see that
the decay rate can be significantly suppressed if enough number of repeaters
are installed. On the other hand, if there are two or more elements that
have been found to be lost and the repeater can ask the previous repeater to
resend the data again, then the classical state will never decay.

\section{Discussion}

\label{sec:dis}

Now the game is over. However, We find ourselves in an embarrassed
situation. Does such an issue still belong to physics? The answer seems to
be YES since there is no any interferences from living beings and the
``repeater"-like device needs energy only. Nevertheless, a digitized state
and its decay clearly neither belong to quantum mechanics nor Newtonian
mechanics. Such a puzzle keeps a challenge for us and is not able to be
answered easily in physical science.

On the other hand, the classical Zeno effect is very popular phenomenon in
the world. Besides the case of Super Mario's prison break \cite{SJGuMario},
the classical Zeno effect exists in many fields, such as the medical field
and the economical field. Take the former as an example: A human being
usually is in a metastable state. Some small deceases can introduce some
fluctuations around the equilibrium point. In these cases, the body's immune
response and a suitable cure will recover the metastable. However, there
exists a class of vital deceases, such as cancer, that will finally drag the
state of human beings from the metastable point and never back if the state
is far from the equilibrium point. Necessary cures clearly can slow down or
ever stop such a process. However, to stop such a process, the most
important is to find and cure the decease as early as possible, then the
probability to recover the metastable state is very high.

\section{Summary}

\label{sec:sum}

In summary, we touched such a possibility of classical Zeno effect in
classical stochastic processes, as illustrated by a scenario of transferring
a classical state through a noise channel. We show explicitly that the Zeno
effect is no longer a patent of quantum mechanics, but exists in classical
stochastic processes. We gave an example of transferring a BMP file through
a noise channel, as simulated by classical Monte Carlo method. The example
manifests that the decay rate of the BMP file will be significantly
suppressed, even never decay, if there are enough number of repeaters.

This work is supported by the Earmarked Grant Research from the Research
Grants Council of HKSAR, China (Project No. CUHK 400807).


\begin{thebibliography}{9}
\bibitem{BMisra77} B. Misra and E. C. G. Sudarshan, J. Math. Phys. \textbf{18%
}, 756 (1977).

\bibitem{KKoshinoPhysRep} K. Koshino and A. Shimizu, Phys. Rep. \textbf{412}%
, 191 (2005).

\bibitem{SJGuMario} S. J. Gu, EPL \textbf{88}, 20007 (2009).
\end{thebibliography}
\end{document}